\documentclass[conference]{IEEEtran}
\usepackage{cite}
\ifCLASSINFOpdf
  \usepackage[pdftex]{graphicx}
  \graphicspath{{./pdf/}{./gfx/}}
  \DeclareGraphicsExtensions{.pdf,.jpeg,.png}
\else
  \usepackage[dvips]{graphicx}
  \graphicspath{{.}{./eps/}}
  \DeclareGraphicsExtensions{.eps}
\fi

\usepackage[caption=false,font=footnotesize]{subfig}

\newcommand{\is}{250pt} %280pt one-column, 210pt two-column

\newcommand{\ist}{230pt}

\begin{document}
% paper title can use linebreaks \\ within to get better formatting as desired
\title{Advanced Programming Platform for efficient use of Data Parallel Hardware}

% use a multiple column layout for up to three different affiliations
\author{
\IEEEauthorblockN{Luis Cabellos}
\IEEEauthorblockA{Institute of Physics of Cantabria (IFCA), CSIC-UC\\
Santander, 39005, Spain\\
Email: cabellos@ifca.unican.es}
%\IEEEauthorblockN{Homer Simpson}
%\IEEEauthorblockA{Twentieth Century Fox\\
%\and
%\IEEEauthorblockN{James Kirk\\ and Montgomery Scott}
%\IEEEauthorblockA{Starfleet Academy\\
%San Francisco, California 96678-2391\\
%Fax: (888) 555--1212}
}


%Telephone: (800) 555--1212, Fax: (888) 555--1212}

% \IEEEspecialpapernotice{(Invited Paper)}

\maketitle

\begin{abstract}
Graphics processing units (GPU) had evolved from a specialized hardware capable
to render high quality graphics in games to a commodity hardware for effective
processing blocks of data in a parallel schema. This evolution is particularly
interesting for scientific groups, which traditionally use mainly CPU as a work
horse, and now can profit of the arrival of GPU hardware to HPC clusters.  This
new GPU hardware promises a boost in peak performance, but it is not trivial to
use.  In this article a programming platform designed to promote a direct use of
this specialized hardware is presented.  This platform includes a visual editor
of parallel data flows and it is oriented to the execution in distributed
clusters with GPUs.  Examples of application in two characteristic problems,
Fast Fourier Transform and Image Compression, are also shown.
\end{abstract}


% \fi
\IEEEpeerreviewmaketitle

% INTRODUCTION %%%%%%%%%%%%%%%%%%%%%%%%%%%%%%%%%%%%%%%%%%%%%%%%%%%%%%%%%%%%%%%%%
\section{Introduction}

%% shaders, y como ha supuesto una revolucion sin saber programar

The game industry saw in the 2000s the revolution of the programmable
shaders. Programmable shaders insert specific code in the 3D graphics pipeline
in order to have customized effects \footnote{Effects not available in the
  graphics hardware.}.  Initially programmable shaders allowed to change the
pixels color, but soon they evolved to be able to modify also the geometry, and
currently they have the flexibility to interact with almost all graphical
elements, achieving the so called General-Purpose computing on graphics
processing units (GPGPU).

The use of GPGPU allow developers to program general processing applications
using graphics hardware. The game companies soon noticed the problem with the
programmable shaders: the graphics are the objective of visual artists, but
programming is a hard and time consuming process that is not taught in art
academies.  And programmers do not have the experience, nor the knowledge to get
the best visual result programming customized effects with graphics hardware.
The solution reached in the game industry was to develop powerful tools, easy to
use for the artist but flexible enough to get all the results that a programmer
can implement directly.  These tools are based on a data-flow programming design
that allows the visual artist to create special effects and display them
on-screen during editing, exactly as they will appear along game execution.  The
tools include the possibility of directed edition showing immediately the
effects of a proposed change, and also the intermediate state of data (as
graphics primitives) in the workflow and so to understand the result of the
change as a whole product or as a sum of transformations.

%% dificultad de uso de GPGPU / Skema como union de los conceptos anteriores

Scientific communities have increasing needs for processing large amounts of
data as fast as possible\cite{DataDeluge}.  Under this demand, they see GPGPU
hardware with its high peak processing power as a valuable resource to handle
for their workflows \cite{Fan:2004:GCH:1048933.1049991}, but they are now in a
similar situation to the the game industry with the programmable shaders: it is
not trivial to program GPU resources for use in real applications.  Efficient
use of GPUs is difficult because although it is possible to use them in almost
every kind of algorithms, only a few of them are executed more efficiently that
in a CPU with a more general architecture.  The input data needs to be sent from
the CPU to the GPU and results returned back, but there is a limited memory
bandwidth between both processors.  Also the internal bus in a GPU is more
powerful, but it has a different organization of memory caches and locations
compared to a CPU, and programs need to take this organization into account to
benefit.  Also the GPU has the advantage of large parallelism thanks to hardware
replication, but it has limited and strict pipelines limiting context switching
between tasks without suffering performance degradation.

%% que es flow-based programming

Data-flow programming is a paradigm that constructs applications as directed
graphs\cite{Arvind:1986:DA:17814.17824}\cite{DBLP:journals/computer/DavisK82}.
The vertex of the graphs are processes and the edges between vertexes define the
input/output of such processes and the path the data should travel.  The
applications in this paradigm are defined changing the set of vertexes and
creating the network of edges between them.  Starting from a clever definition
of vertexes and changing only the edges, it is possible to create many different
applications under this paradigm.

%% comentar el scientific workflow

The data-flow programming paradigm is close to the preferred work methodology
used by scientific communities.  They usually share the data from experimental
sources and define scientific workflows to analyze that data.  Scientific groups
with better computing skills also share computational services to analyze data.
However researchers do not usually apply the data-flow paradigm directly to
program their applications, they rather compose them through successive filter
and processing steps starting from the raw data obtained from the experimental
setup.

However, there are several solutions to compose workflows in a scientific
programming context, starting with the popular and powerful LabView
software~\cite{LabView}, the Kepler System~\cite{Ludäscher05scientificworkflow},
or others adapted to a Grid computational infrastructure\cite{citeulike:798241}.

The proposal presented here aims to use the data-flow paradigm starting at the
basic level, constructing modules from a well defined set of processes,
conceived as orthogonal components, including data parallelism, and that
communicate between them to build applications.

This Distributed Programming Platform for Data Parallel Algorithms will also
benefit of powerful visual tools like those already developed under the
data-flow paradigm~\cite{Johnston:2004:ADP:1013208.1013209}.  These tools will
include a user friendly editor to program data-flow applications when the
algorithm fits into a data parallelism model, and also a service able to execute
the resultant program flows in the most efficient way using computers with one
or more GPUs.

The architecture explained in this paper is designed to allow the user to build
the flow once and be able to execute it with different data sets and on
different distributed computing hardware offering GPU resources.

% COMPUTATIONS %%%%%%%%%%%%%%%%%%%%%%%%%%%%%%%%%%%%%%%%%%%%%%%%%%%%%%%%%%%%%%%%%
\section{Implementation}

\subsection{GPU Framework}

%% Seleccion de OpenCL y data parallelism model
The first step in the development of a Data-Parallel Platform is the selection
of a GPGPU platform.  Currently there are are two major platforms available:
CUDA and OpenCL\cite{gpgpu.org}.  CUDA stands for Compute Unified Device
Architecture, and it is a computing engine developed by Nvidia Corporation
enabling access to Nvidia GPUs as a GPGPU platform\cite{cuda2.0}.  OpenCL is an
specification made by the Khronos Group, of an open standard for general purpose
parallel programming\cite{opencl1.0}.  Although both platforms are presented to
use GPU hardware as GPGPU platforms, they also support manycore and multicore
hardware, so they are able to execute code in both CPUs and GPUs, provided the
corresponding driver.

CUDA offers higher quality libraries and also includes better development tools,
like a debugger and an emulator.  Applications in CUDA require less setup code,
and the performance compares favorably with
OpenCL\cite{DBLP:journals/corr/abs-1005-2581}.

Both platforms use the C language as their reference model, and have similar
memory and concurrency characteristics, so converting programs between both
platforms is not difficult.

On the other hand, OpenCL has a clear advantage over CUDA: while CUDA is
designed to work with NVIDIA hardware, OpenCL, as an open standard, has already
drivers implemented for NVIDIA, ATI and Intel hardware, both for GPUs and CPUs.

From the point of view of a potential user of the Data-Parallel Platform, this
is the most relevant argument. In fact, potential users of the
Data-Parallel-Platform are not interested in the tools that the developers will
use, nor in the setup details, but are critically concerned about availability
of the platform for their existing hardware.  Based on these arguments, OpenCL
was selected as the platform to build the Data-Parallel Platform.

The Data-Parallel Platform uses OpenCL in two different ways.  Firstly, the
OpenCL software development kit is used to execute the program flows in the GPU
hardware, including manycore and multicore hardware when available

% \footnote{An advantage that we'll have with the use of OpenCL (also CUDA) in hardware
% isolation.}).

Secondly, the OpenCL C programming language is employed to codify the behavior
inside the vertexes of the Data-Parallel Program graphs.  As it will be shown
below, there is a direct translation between the Data-Parallel Platform vertexes
and OpenCL C source code. A simple example can be seen comparing middle and
bottom sections in Table~\ref{tab:compare}.

\begin{table}[t]
\begin{center}
\begin{tabular}{|p{\ist}|}
\hline\scriptsize
\begin{verbatim}
for( int i = 0 ; i < MAX ; i++ ){
    z[i]=x[i]+y[i];
}
\end{verbatim}
\\ \hline\scriptsize
\begin{verbatim}
__kernel adder( global float * x, global float * y, 
                global float * z ){
    int i = get_global_id(0);
    z[i]=x[i]+y[i];
}
\end{verbatim}
\\ \hline\scriptsize
\begin{verbatim}
"adder":{
  "body":
    "int i = get_global_id(0);\nz[i]=x[i]+y[i];\n",
  "io":{
    "x":{ "data":"float", "type":"InputPoint"},
    "y":{ "data":"float", "type":"InputPoint"},
    "z":{ "data":"float", "type":"OutputPoint"}}
}
\end{verbatim}
\\ \hline
\end{tabular}
\end{center}
\caption{Comparative of the implementation of a loop in the different platforms.
  Top: the loop in C-like pseudocode. Middle: the loop using an OpenCL kernel.
  Bottom: the loop in Data-Parallel JSON format.}
\label{tab:compare}
\end{table}

This strategy can be seen as a complexity reduction of the access to GPU
programming\cite{Ueng:2008:CRG:1485701.1485702}: hand coding complexity for the
final user is much reduced as OpenCL functions input and output parameters are
limited. However it implies also a limitation on the complexity of the problem
being coded, and addressed.

OpenCL support two execution models, data parallel and task parallel programming
models.  The Data-Parallel Platform will use only the data parallel programming
model of OpenCL to offer data-flow programming to users.  In this data parallel
model, a sequence of instructions is applied to multiple elements in memory.
Each one of these elements is called a work-item and the parallelism is achieved
executing the sequence of instructions at the same time over all the work-items.
In the Data-Parallel Platform a one-to-one bind between the work-item in memory
and the kernel currently executed is established.  In this way the input
data-flow in a Data-Parallel program is split into chunks of work-items, then
executed in parallel using OpenCL and finally the result is re-joined to compose
the output data-flow.

\subsection{The Data Parallel Model}
%% uso de DAG

The use of the data parallel model in OpenCL and the division in blocks of
work-items requires that the Data-Parallel Programs are strictly Directed
Acyclic Graphs (DAGs). This requirement avoids return edges, that would
complicate the parallelism of blocks of elements if a vertex would have to wait
for the output of a posterior vertex.
  
%% desarrollo en dos capas principales

\begin{figure}[t]
\begin{center}
\includegraphics[width=\is]{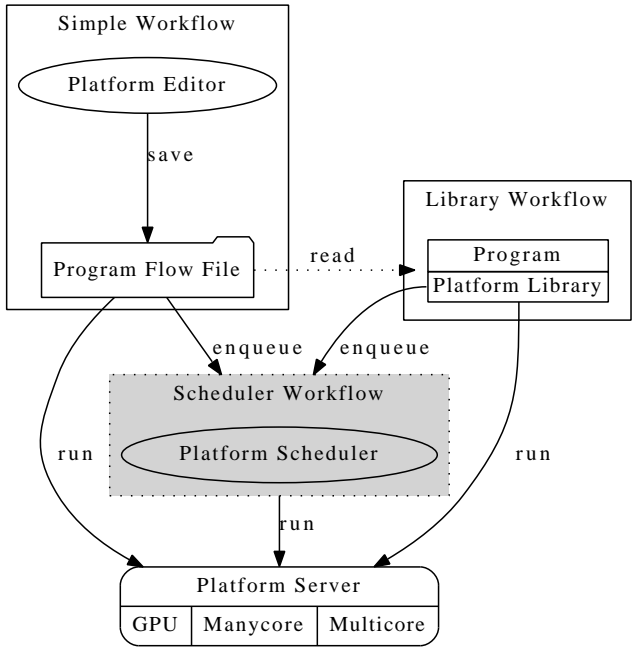}
\end{center}
\caption{Data-Parallel Platform workflow scheme. The graph shows the two
  different ways to use the Data-Parallel Platform: 1) As a library from a user
  application; 2) from the Data-Parallel Platform Editor, either running
  directly in the Data-Parallel Platform Server, or, in the future, as a job
  system.}
\label{fig:workflow01}
\end{figure}

The Data-Parallel Platform is structured as several layers or components in
order to allow the execution of Data-Parallel programs in different ways: direct
execution, scheduled execution on a queue, or integrated in existing
applications using a library.  These different possibilities are represented in
figure~\ref{fig:workflow01}.

The most basic example would start with the creation of a Data-Parallel Program
using the Data-Parallel Editor (see below), then selecting the input files to be
processed, and finally executing the program in a Data-Parallel Server (also
presented later).

The same Data-Parallel Program created using the editor can be executed by a
program using the functions from the Data-Parallel Platform library.  Execution
using a queue system is being implemented under a wider scope, in a Distributed
Data-Parallel Platform including a Data-Parallel Scheduler acting as a batch
system for Data-Parallel Programs\footnote{The whole set of tools described here
  is being further developed under the name of the Skema Platform.}.

\subsection{A Visual Editor of Data-Parallel Programs}

A Data-Parallel Program Editor has been implemented as a visual tool following
the Blender\footnote{Blender is the free open source 3D content creation suite.}
Compositor style.  The edition of a complete Data-Parallel Program has two
parts.  The first one is the definition of the nodes: individual nodes are
created and can be modified individually, including its input/output set and its
body main program in OpenCL C Programming Language.  The second part is the
definition of the data flow between instances of the nodes. Nodes must be
instantiated and arranged into a processing network.

\begin{figure}[t]
\begin{center}
\includegraphics[width=\is]{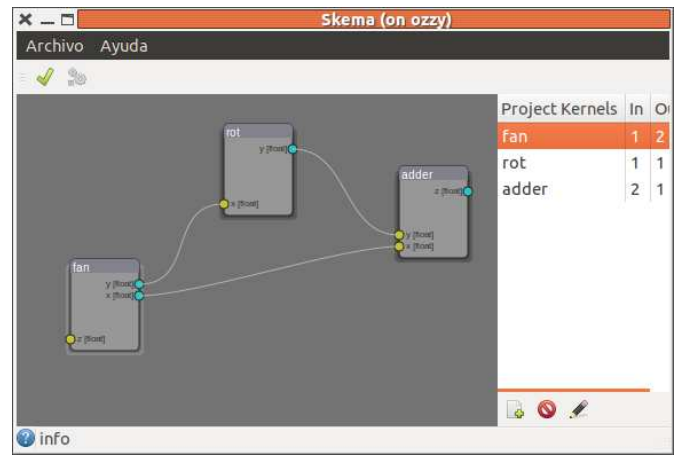}
\end{center}
\caption{The Visual Editor showing a basic Data-Parallel Program. The data flows
  from left to right with a floating number as input and a floating number as
  result.}
\label{fig:program01}
\end{figure}

Figure~\ref{fig:program01} shows how a Data-Parallel Program appears in the
visual editor.

%% componentes de un flujo de programa en skema

\begin{figure}[t]
\begin{center}
\includegraphics[width=\is]{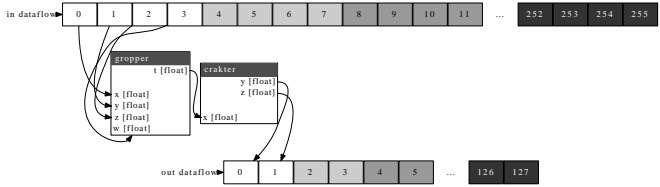}
\end{center}
\caption{Representation of the execution of a Data-Parallel Program. The
  Data-Parallel Program get chunks of data from an input stream, executes the
  programming code included in the nodes in parallel for each of the elements of
  that chunk, and generates an output stream composed of the results re-joined
  in adequate order.}
\label{fig:stream01}
\end{figure}

A Data-Parallel Program is a data-flow application created to be executed in the
Data-Parallel Platform. Its main components are the following ones:

\begin{LaTeXdescription}
\item[Type] The available data types in the Data-Parallel Platform: OpenCL
  1.0~\cite{opencl1.0} data types are used, including scalar and vector data
  types.

\item[Input/Output Point] Points attached to vertexes in the Data-Parallel
  Programs. The set of points of a Node define the possible communication
  channels between instances of that node.

\item[Node] A node defines the behavior of the graph vertexes in a Data-Parallel
  Program. It is composed of a set of Input/Output points (at least one of each
  type) and a main program body coded using the OpenCL C Programming Language
  specification.

\item[Instance] An instance of a node is a vertex in the Data-Parallel
  Program. In the example shown in the previous figure~\ref{fig:program01} 
  there are three instances of three different nodes.

\item[Arrow] An arrow is an edge between two instances or
  vertexes. Specifically, an arrow connects an output point of an instance with
  a compatible input point from a different instance. The points are compatible
  if they have the same base scalar type\footnote{The base type of an scalar is
    that scalar data type. The base type of a vector data type is the scalar
    element of the vector}. A point from an instance without a connected arrow
  is named an unassigned point or a free point.

\item[Program] A Program is the directed acyclic graph of instances and arrows
  than can be executed in the Data-Parallel Platform. The diagram shown in
  figure~\ref{fig:program01} shows a basic Data-Parallel Program.

\item[Stream] A stream is a continuous flow of data with a defined type and
  related to a free point of a Data-Parallel Program. The execution of a program
  requires one or more input streams and one or more output streams. The
  figure~\ref{fig:stream01} describes the execution of a Data-Parallel program
  over an input stream to generate an output stream.

\item[Data-flow] The Data-flow is the set including the input stream and the
  output stream in a Data-Parallel Program execution.

\end{LaTeXdescription}

Continuing with the editing process, once the nodes, instances and arrows
between instances of a Data-Parallel Program are defined, all the corresponding
information is exported to a JSON\cite{rfc4627} file.  This file stores this
information in a format used in the execution of Data-Parallel programs, either
for sending it to a Data-Parallel Server or for connecting to the Data-Parallel
Platform Library.  The Table~\ref{tab:simple} shows a basic example of exported
information corresponding to the Data-Parallel Program described in
Figure~\ref{fig:program01}.  The corresponding JSON file is used by the
Data-Parallel Platform to execute the program flow.

\begin{table}[t]
\begin{center}
\begin{tabular}{|p{\ist}|}
\hline\scriptsize
\begin{verbatim}
"kernels":{
  "adder":{
    "body":"int i=get_global_id(0);
            z[id]=x[i]+y[i];",
    "io":{
      "x":{"data":"float","type":"InputPoint"},
      "y":{"data":"float","type":"InputPoint"},
      "z":{"data":"float","type":"OutputPoint"}}},
  "fan":{
    "body":"int i=get_global_id(0);
            x[i]=z[i].x;
            y[i]=z[i].y;",
    "io":{
      "x":{"data":"float","type":"OutputPoint"},
      "y":{"data":"float","type":"OutputPoint"},
      "z":{"data":"float2","type":"InputPoint"}}},
  "rot":{
    "body":"int i=get_global_id(0);\ny[i]=x[i]<<16;",
    "io":{
      "x":{"data":"float","type":"InputPoint"},
      "y":{"data":"float","type":"OutputPoint"}}}},
"nodes":[[0,{"kernel":"fan"}],
  [1,{"kernel":"rot"}],
  [2,{ "kernel":"adder"}]],
"arrows":[{"output":[0,"x"],"input":[2,"x"]},
        {"output":[1,"y"],"input":[2,"y"]},
        {"output":[0,"y"],"input":[1,"x"]}]
\end{verbatim}
\\ \hline
\end{tabular}
\end{center}
\caption{JSON format corresponding to a Data-Parallel program. This basic
  example corresponds to the program previously shown in
  Figure~\ref{fig:program01}, composed of three nodes and three instances, and
  the corresponding data-flow between instances composed of three edges.}
\label{tab:simple}
\end{table}

\subsection{The Data-Parallel Server}

%% estructura del servidor de skema

The Data-Parallel Server is the module in the platform that executes the
Data-Parallel programs on an input data-flow to obtain an output data-flow.  For
that reason it is the only module that actually requires the OpenCL driver and
also direct access to the associated hardware.  The server is in charge of
communicating the state of the OpenCL platform, the state of the GPGPU hardware
and its characteristics, and also the running progress of Data-Parallel
programs.  It uses a simplified approach for external communication based on
REST (Representational State Transfer)\cite{Fielding:2002:PDM:514183.514185}
with HTTP networking protocol and JSON documents as information exchange format.
The Data-Parallel Server currently is not RESTful as not all the REST
architectural elements are implemented, in particular the layered and cacheable
properties, although it is planned to include these features when possible in
future developments to improve its scalability.

\begin{figure}[t]
\begin{center}
\includegraphics[width=\is]{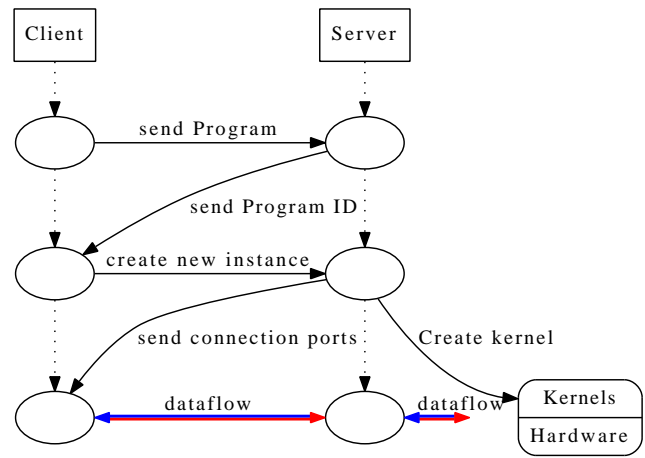}
\end{center}
\caption{Distributed Run Protocol. Client starts a running instance in the
  server, to execute the desired program, using a web API in http with JSON as
  data language; once the instance is created, the client sends and receives
  data from the server via the TCP protocol.}
\label{fig:runprotocol01}
\end{figure}

The Data-Parallel Server executes the Data-Parallel Programs using a simple Run
Protocol to connect with the clients.  As shown in
Figure~\ref{fig:runprotocol01}, this protocol defines the order of the steps to
execute a program over a data-flow: send the Data-Parallel Program to the
server, initialize the execution of the program and finally send the input
data-flow and receive back the output data-flow.  It is important to notice that
the first step could be skipped if the program is transferred previously.  For
this purpose, an unique ID can be associated with the JSON representation of the
program, a program ID.  This option may save a significant time if the same
Data-Parallel Program is to be executed with different input streams.

% RESULTS %%%%%%%%%%%%%%%%%%%%%%%%%%%%%%%%%%%%%%%%%%%%%%%%%%%%%%%%%%%%%%%%%%%%%%
\section{Examples}

In order to show the capabilities of our Data-Parallel Platform two working
examples are presented below. The first example computes the discrete Fourier
transform using the Cooley-Tukey algorithm. The second example is a simple lossy
image compression application using a visual vector quantization of blocks.

Both examples were tested in a server node running the Data-Parallel Server and
a desktop computer running the client programs.  The server node was a Megware
Computer solution equipped with an Intel Xeon X5550 2.66GHz Quad Core processor
and 4GB memory. It had four NVIDIA Tesla C1060 GPUs installed.
 
The desktop computer was an HP Proliant ML330 G6. Both computers are
interconnected using a Gigabit Ethernet LAN.

\subsection{Discrete Fourier Transform Example}

The discrete Fourier transform (DFT) is a mathematical transform of a signal
between discrete domains used for Fourier analysis. The DFT is widely used in
signal processing, to analyze the frequencies of a signal, data compression
eliminating frequencies with less information in a signal, polynomial
multiplication and convolutions. All these applications depend upon an efficient
calculation of this transformation, so it is a good example to test the speedup
of this DFT using the Data-Parallel Platform on GPU hardware.

The Cooley-Tukey\cite{CooleyTukey} algorithm was used to implement the Fast
Fourier Transform.  This algorithm, one of the most used in DFT, recursively
calculates a DFT of N elements using two DFT of sizes $N1$ and $N2$ having
$N=N1\cdot N2$. When $N$ is highly composite \footnote{Highly composite numbers
  are numbers which factors completely into small prime numbers.} the DFT
computation time can be reduced from $O(N^2)$ to $O(N \cdot log N)$.  In
particular, the radix-2 Cooley-Tukey algorithm was used, where the decimation of
the DFT is done with two interleaved DFT of size $N/2$ in each recursive step.

Using this radix-2 decimation, the computation of the last $k$ steps can be sent
to the Data-Parallel Platform Server, parallelizing the calculus of a large
number of DFT of size $2^k$.  The $2^k$ DFT is computed in a simple Program Node
and the DFT flow is the input flow of the Data-Parallel Program.

\begin{figure}[t]
\begin{center}
\includegraphics[width=\is]{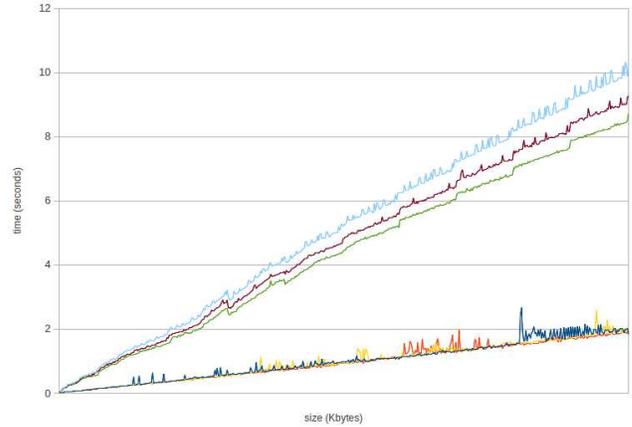}
\end{center}
\caption{Comparing DFT on CPU versus GPU execution. The size of the data used
  span from 20 Kbytes to 10 MBytes. Three FFT sizes 2, 4 and 8 values in both
  types of tests (CPU and GPU) are used. All GPU tests remain well below 2
  seconds in execution time while the CPU tests increase up to $8 \sim\ 10$
  seconds.}
\label{fig:fft01}
\end{figure}

Figure~\ref{fig:fft01} presents the result of executing the Data-Parallel
Program with a flow of DFT with sizes 2, 4 and 8 and the execution of the same
data with a CPU implementation of the Cooley-Tukey algorithm for the same
sizes. The time increase is linear in both implementations, as it should be, but
it can be seen that the Data-Parallel Program is approximately five times
faster, although it has to send the data over the local network to the
Data-Parallel Server.

The Data-Parallel Program has also a remarkable advantage over the full CPU
implementation. While the Data Parallel platform is executing the DFT
calculations, the CPU usage in the local computer is only around $10\%$ and
corresponding to I/O time; in contrast the CPU implementation requires $\sim
90\%$ CPU usage.

\subsection{A second example: Image Block Compression}

\begin{figure*}[!t]
\centerline{
\subfloat[Original]{
  \includegraphics[width=2.0in]{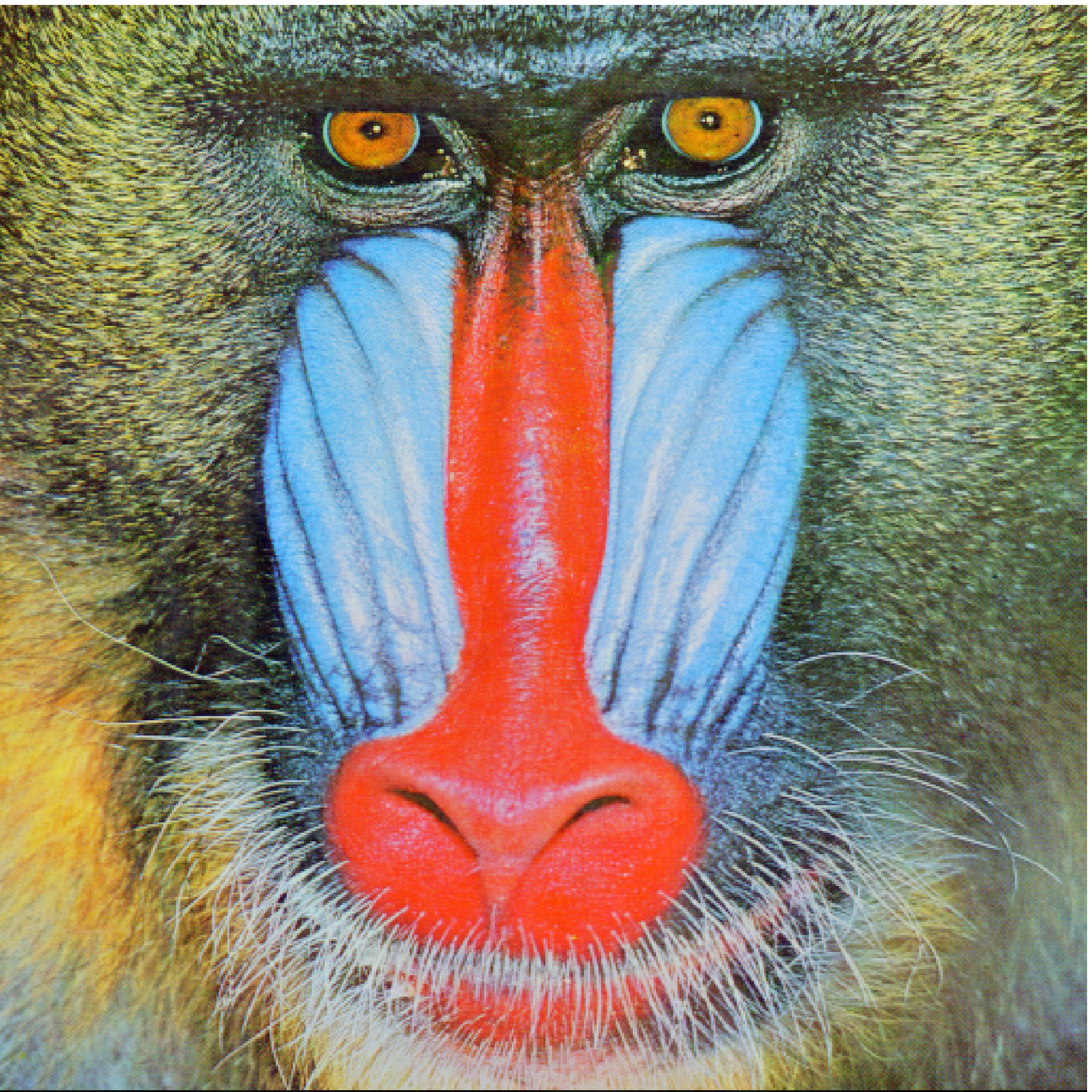}
  \label{fig:sub01:compress}
}
\hfil
\subfloat[Luminance compressed]{
  \includegraphics[width=2.0in]{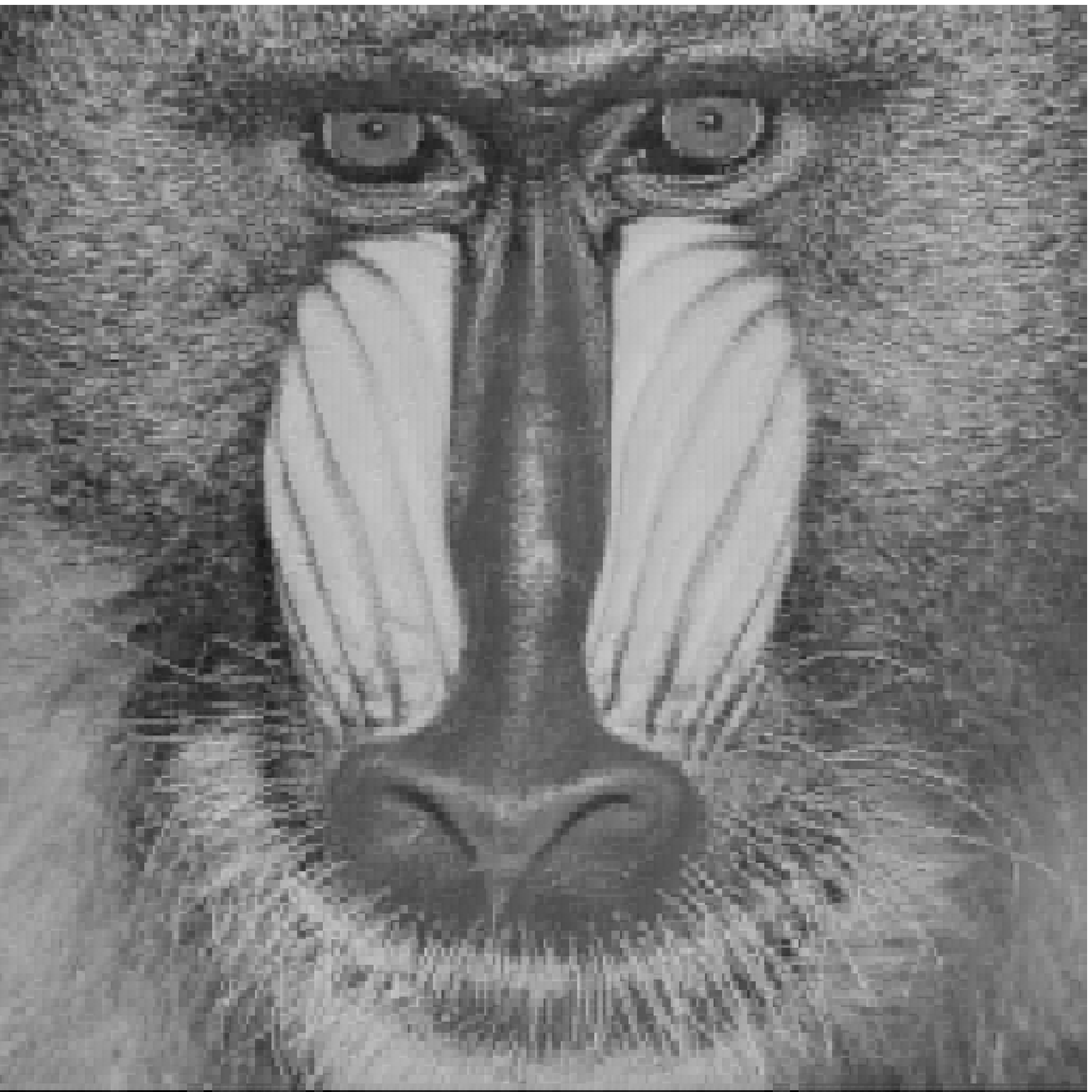}
  \label{fig:sub02:compress}
}
\hfil
\subfloat[Result]{
  \includegraphics[width=2.0in]{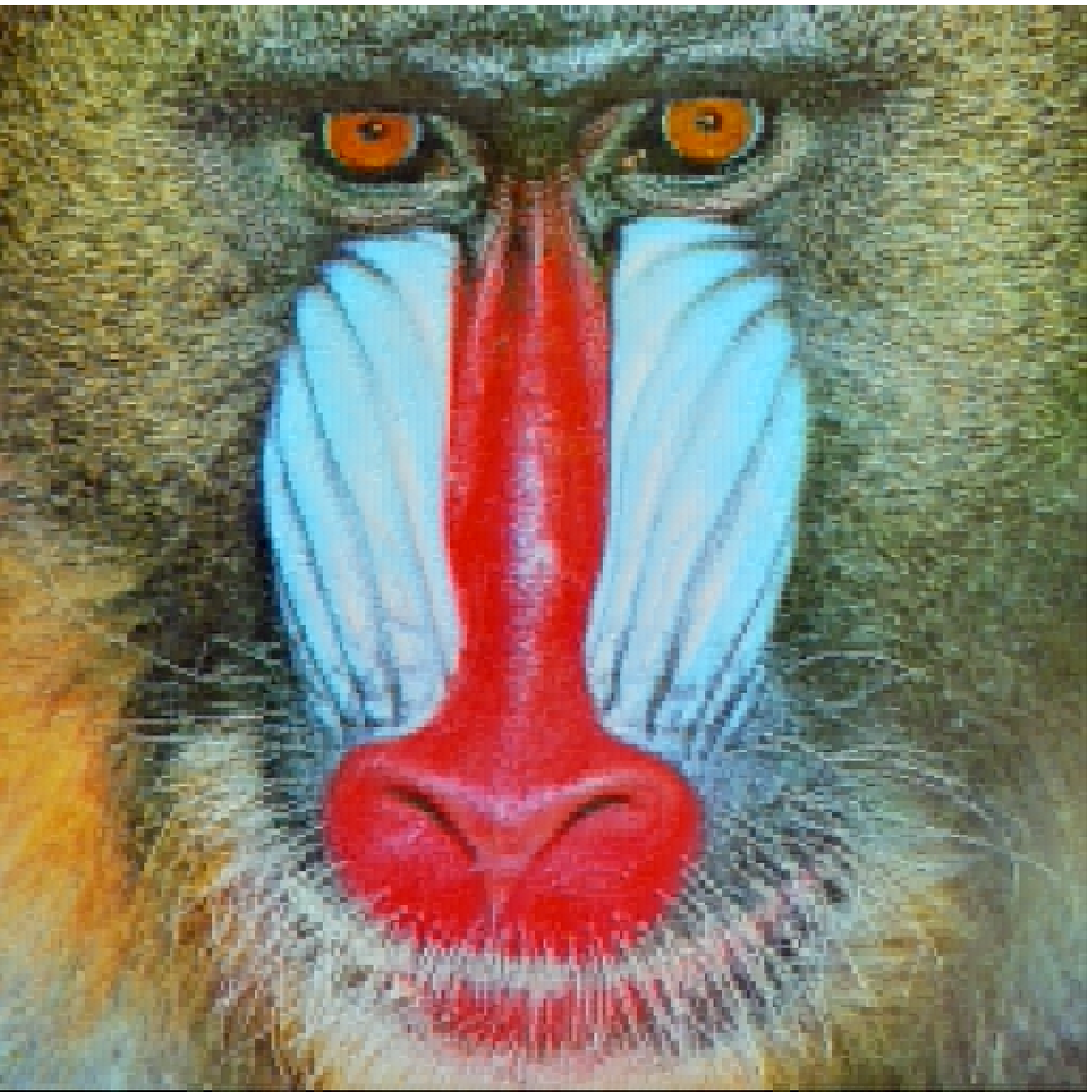}
  \label{fig:sub03:compress}
}}
\caption{Image compression from (a) original image to (c) compressed result. The
  block compression is done in the (b) luminance of the image, the colour layer
  is scaled down 1/4th from the full picture. Uncompressed size is $\sim
  770$ Kbytes, and compressed size is $\sim 80$ Kbytes}
\label{fig:compress}
\end{figure*}

The purpose of image compression is to represent images using less data in order
to save storage costs or transmission time.  A raw image, without compression,
can be quite large, usually up to several megabytes, and compression can reduce
the file size very significantly. The image data can be compressed in such way
that the exact original data to can be recovered from the compressed data, or
losing information in the image data but reaching better compression rates.

Lossy compression is usually based on techniques that remove details that humans
do not notice. In this example a lossy image compression has been implemented
using the Data-Parallel platform and employing well known methods.  The first
method used is the conversion of red-green-blue image data to a chrome-luminance
representation, followed by color sub-sampling to scale 1/4 the from the full
size using the fact that the human eye is more sensitive towards light intensity
variation than color variation \footnote{Nobody will notice the colour
  downscale}. The second method applied, explained in references
\cite{citeulike:5759381} \cite{qiu1995}, divides the image luminance in blocks
of $4x4$ pixels, and determines a code book of N representative mean blocks.
With this code book image blocks are encoded, using intensity deviation, instead
of using the full information of the 16 pixels.

The implemented algorithm uses the following five steps:

\begin{enumerate}
\item \label{itm:a}Convert to Chrome + Luminance representation
\item \label{itm:b}Downscale Chrome layer
\item \label{itm:c}Calculate Directional Derivative of Luminance
\item \label{itm:d}Apply k-means for calculate codebook
\item \label{itm:e}Compress Luminance in blocks
\end{enumerate}

Steps \ref{itm:a}, \ref{itm:b} and \ref{itm:c} are calculated in the
Data-Parallel Platform, while the creation of the code book (step
\ref{itm:d}) is made in the CPU with the returned data.  Once the code book is
created, the information is sent back to the GPU to calculate the compressed
data.

The compression is noticeable, as can be seen in Figure~\ref{fig:compress}, both
in quality and in the reduction factor, but the objective with this example is to show how
the Data-Parallel Platform can be used to build a working application.  The
sections of the program executed in the GPU were created using the visual
editor, and during the execution of the image compression this work was
distributed on a running Data-Parallel Server using the Data-Parallel Library.

% CONCLUSION%%%%%%%%%%%%%%%%%%%%%%%%%%%%%%%%%%%%%%%%%%%%%%%%%%%%%%%%%%%%%%%%%%%%
\section{Conclusions and Outlook}

A Data-Parallel Platform has been designed supporting the use of GPGPU on
clusters allowing to access to the power of GPUs as a service, with the
advantages this means for the implementation of work-flows and schedulers.
Difficulties of GPGPU programming are reduced thanks to a clear programming
model using the OpenCL platform and modeling the problems using DAGs, and
offering a visual editor tool for final users powerful enough to exploit the
GPGPU syntax.

%% la construccion de supercomputadores actualmente se basa en GPU

There is an increasing trend to use GPUs specialized processors as common
building blocks of supercomputers.  China's Tianhe-1A supercomputer achieved in
October 2010 the number one in the TOP500 ranking using graphics chips, and in
the march 2011 3 of the top 5 supercomputers\cite{top500:2011:06} were using
mixed architectures with both CPUs and GPUs.  Although the increase in
performance thanks to the use of GPUs seems very high, with up to a 20x factor,
GPUs require specialized programing, and the lack of advanced programming tools
and languages with limited features is a problem\cite{varhol:de:2010}.

%% mejoras de toda la plataforma
%% usar Graph theory para mejorar
The Data-Parallel Platform presented does not aim to be the best tool for
performance, and it's not yet fully completed to offer all the characteristics
planned, but it is a solution prepared to allow distributed computing with GPGPU
hardware.

Many improvements will be required to make it a production tool.  For example,
regarding performance of program executions, the gap when using a cascade of
instances due to inefficient movement of data between them, has to be solved

Graph theory must be revisited in order to further optimize the Data-Parallel
Programs.  In particular to understand how to split a Data-Parallel Program into
several concurrent flows.  There is also the possibility of include
characteristics of other distributed solutions in the Data-Parallel Platform,
like high availability, large scalability or Map/Reduce technologies.

Also the design of a job system for Data-Parallel Programs running on
Data-Parallel servers, as part of a Distributed Data-Parallel Platform, will
allow a better scale of applications and a better use of GPGPU resources,
especially in computer clusters with GPU hardware.

\section*{Acknowledgment}

This works was supported by the Ministry of Science and Innovation of Spain and
their National Scientific Research, Development and Technological Innovation
Plan (National R\&D\&i Plan) at the University of Cantabria.

\IEEEtriggeratref{15}
\bibliography{IEEEabrv,../refs/bibtex}{}
\bibliographystyle{IEEEtran}

\end{document}